\documentclass[runningheads]{llncs}
\usepackage{graphicx}
\usepackage[square,numbers]{natbib}
\usepackage{enumitem}
\usepackage[dvipsnames]{xcolor}

\begin{document}
\title{Requirements for Explainability and Acceptance of Artificial Intelligence in Collaborative Work}
\titlerunning{Requirements for AI in Collaborative Work}
%
\author{Sabine Theis\inst{1}\orcidID{0000-0002-3422-3734}\and{Sophie Jentzsch}\inst{1}\orcidID{0000-0001-6217-8814}\and{Fotini Deligiannaki}\inst{2}\orcidID{0000-0002-9479-8767}\and{Charles Berro}\inst{2}\orcidID{0000-0002-2662-8774}\and{Arne Peter Raulf}\inst{2}\orcidID{0009-0003-8672-3014}\and{Carmen Bruder}\inst{3}\orcidID{0000-0003-2638-2361}}

\authorrunning{S. Theis et al.}
%
%
\institute{Institute for Software Technology, Linder H\"ohe, 51147 Cologne\\
%
\and Institute for AI Safety and Security, Rathausallee 12, 53757 Sankt Augustin\\
%
\and Institute for Aerospace Medicine, Sportallee 5a, 22335 Hamburg\\
\email{\{firstname\}.\{lastname\}@DLR.de}} 
%
%


\maketitle              
\begin{abstract}
The increasing prevalence of Artificial Intelligence (AI) in safety-critical contexts such as air-traffic control leads to systems that are practical and efficient, and to some extent explainable to humans to be trusted and accepted. The present structured literature analysis examines $n = 236$ articles on the requirements for the explainability and acceptance of AI. Results include a comprehensive review of $n = 48$ articles on information people need to perceive an AI as explainable, the information needed to accept an AI, and representation and interaction methods promoting trust in an AI. Results indicate that the two main groups of users are developers who require information about the internal operations of the model and end users who require information about AI results or behavior. Users’ information needs vary in specificity, complexity, and urgency and must consider context, domain knowledge, and the user’s cognitive resources. The acceptance of AI systems depends on information about the system’s functions and performance, privacy and ethical considerations, as well as goal-supporting information tailored to individual preferences and information to establish trust in the system. Information about the system’s limitations and potential failures can increase acceptance and trust. Trusted interaction methods are human-like, including natural language, speech, text, and visual representations such as graphs, charts, and animations. Our results have significant implications for future human-centric AI systems being developed. Thus, they are suitable as input for further application-specific investigations of user needs. 

\end{abstract}

\section{Introduction} \label{s:introduction}
Humans will collaborate with \textit{artificial intelligence (AI)} systems in future living and working environments. In particular, this will characterize aviation, medicine or space travel activities. These outstanding safety-critical application areas require -- more than others -- the consideration of individual requirements of human operators in the design of collaborative assistance systems. In this context, the German Aerospace Center (DLR) is developing guidelines for the human-centered collaboration design between users and AI systems. The focus is on tasks and contexts where operators, such as \textit{air traffic controllers (ATCOs)}, medical professionals, or operators of space systems work collaboratively with AI to achieve efficient and safe operation. Especially humans as the operators of AI systems form a thematic priority, together with the question of how explainability and acceptance can be assured for the users of AI systems. %
To develop an explainable and acceptable AI pilot and AI air traffic co-controller, this article examines previously noted \textit{user requirements} in collaboration with artificial intelligence. In general, \textit{user requirement analysis} denotes an iterative process in which one identifies~\cite{nuseibeh2000requirements}, specifies, and validates functional and non-functional characteristics of an IT system~\cite{finkelsteiin2000software, kujala2005role} together with individual users and user groups. %
The main goal of this user requirement engineering process is to ensure that the system to be developed meets the needs of its users~\cite{kujala2008effective, bano2013users}. This requires a deep understanding of user characteristics, goals, and tasks, as well as the context in which the software or system will be used~\cite{dalpiaz2020requirements, maalej2019data}. User requirement engineering is a critical part of the software development process, as it can significantly impact the success or failure of the final product~\cite{finkelsteiin2000software}.\\
One variable within the context of requirements engineering of data-driven systems refers to the data and information that a user requires in order to achieve their goals or perform their tasks~\cite{theis2019you, theis2019predicting, sutcliffe1998scenario}. A focus on \textit{users’ information needs (IN)} and seeking behavior during user requirement engineering focuses the development process less on technology and more on the essential result of information technology, namely the transfer of information to the human, which is especially relevant for human-AI interaction or data visualization systems~\cite{cai2019hello, taggart1977survey}. Information needs essentially describe the gap between a current and the desired state of knowledge that needs to be filled to achieve a goal which in the present case is to understand an AI~\cite{wilson1981user}. IN is defined by an individual and can vary in specificity, complexity, and urgency while being induced by social, affective, and cognitive needs.\\
The main contribution of the present article is a brief and understandable overview of the technical background of AI, \textit{explainable artificial intelligence (XAI)}, and \textit{human-in-the-loop (HITL)} in AI development. Through an interdisciplinary synthesis of computer science and psychological knowledge the present article address the need for human-centered explainable and trusted artificial intelligence by eliciting user requirements through a structured analysis of previous work on human-AI interaction.
\section{Background}\label{s:background}
The following paragraphs briefly introduce related technological aspects to establish a common understanding of the broader context. First, challenges and state-of-the-art of XAI are discussed from a technical perspective in Sec.~\ref{ss:background:technical}. Sec.~\ref{ss:background:human} then provides a human perspective on explainability in intelligent systems. Finally, Sec.~\ref{ss:background:HITL} brings both perspectives together by discussing the HITL paradigm. 

\subsection{Technical Perspective}\label{ss:background:technical} 
AI is an umbrella term for a wide variety of different systems and techniques. %
When AI first emerged in the 1950s, the main focus was on symbolic AI, where real-world concepts are represented by symbolic entities, and human behavior is expressed by explicitly formulated logical rules. However, with exponentially growing computational resources and a stronger focus on statistical approaches, AI went through a sharp paradigm shift in the 1990’s: from a logic-based to a data- and representation-driven doctrine \cite{Rumelhart1986LearningRB}. This was the start of the era of \textit{machine learning (ML)} and later \textit{deep learning (DL)}, which is a subdomain of ML using (deep) neural networks~\cite{mitchell2021ai}. ML has reached multiple milestones in various domains~\cite{t5_raffel, nowcast_Ravuri2021, Assael.2022}, exploiting massive amounts of data with sophisticated algorithms. \\
Nowadays, the field of AI is strongly dominated by ML and DL, turning the vast majority of concrete XAI implementations towards ML-based sub-problems. Understanding ML-based systems is especially challenging and relevant for several reasons, outlined in the following paragraphs.

\subsubsection{ML and explainability}
XAI is a constantly growing research field, yet there exists no single established definition of explainability and related concepts such as transparency or interpretability~\cite{arrieta2020explainable}. The EASA defines \emph{explainability} as the “\textit{capability to provide the human with understandable and relevant information on how an AI/ML application is coming to its results.}”~\cite{EASA2021concept}. A strongly related term in literature is \textit{interpretability} as it is often defined similarly, e.g., the users’ ability to “\textit{correctly and efficiently predict the method’s results}”~\cite{kim2016examples}.\\ 
Achieving system explainability is a relevant requirement for numerous reasons, essentially though because ML-based systems’ decisions affect many aspects of our daily lives and need to be proven reliable. %
While empirical evidence on the effects of system explainability on users’ trust remains inconclusive, explainability certainly supports \textit{trustworthiness}~\cite{kastner2021relation}. 
In addition, Gerlings et al. outline a comprehensive list of motivations for XAI which are, among others, generation of trust and transparency, following compliance and regulations (e.g. GDPR), social responsibility and risk avoidance~\cite{gerlings2021reviewing}.\\ 
Although it is evident that XAI is a fundamental requirement it comes along with multiple technical and systemic obstacles. In a standard ML pipeline, enormous amounts of data are fed into an algorithm that autonomously identifies and encodes relevant patterns into a model. %
%
In contrast to symbolic AI, state-of-the-art ML models encode real-world information and human knowledge implicitly into inherently opaque models. %
It is, therefore, especially challenging to retrace their latent reasoning and portraying their insights rationally, which is why they are frequently referred to as \textit{black boxes}. %
%
\begin{figure}[!ht]
\centering
\includegraphics[scale=1.2]{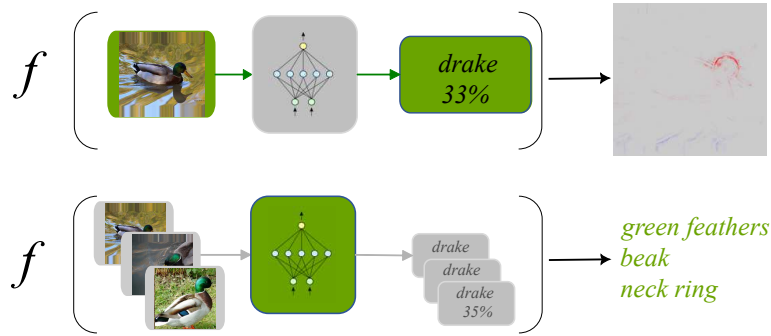}
\caption{The concept of a local (top) and global (bottom) explanation function $f$ is depicted. LRP~\cite{montavon2019layer}, the former explanation method
, solely depends on the specific input/output pair. The output highlights the useful to the network image area (as seen in red, the duck’s head). The later yields abstract features/concepts, targets the AI model in itself and explains what features are used in its decision. The input images are taken from the ImageNet data set~\cite{imagenet}.} \label{fig:local_global_explanations}
\end{figure}
\footnotetext{The following open API was used for generating explanations: https://lrpserver.hhi.fraunhofer.de/image-classification}

\subsubsection{Explanation characteristics}
Technically, explanation approaches can be divided by a row of characteristics~\cite{EASA2021concept,mohseni2021multidisciplinary}: 
\paragraph{Global vs. local} While global explanation approaches seek to describe the overall model, answering the question \textit{what information does the model utilize to answer?}, local approaches are designed to explain specific model outputs or the role of particular input samples, i.e, \textit{why did the model yield a specific output from a specific input?}. Both approaches are illustrated in Fig.~\ref{fig:local_global_explanations}. 
\paragraph{Model-specific vs. model-agnostic} Depending on the specific system or the types of input data, different explanation techniques can be appropriate. Other approaches claim to be model agnostic. That means they can be applied to every type of underlying ML model. 
\paragraph{Intrinsic vs. post-hoc explainability} Some ML models, e.g., linear and logistic regression or decision trees, are intrinsically interpretable. These models can be explained by restricting the models’ complexity~\cite{molnar2020interpretable}. In contrast, models that do not possess that characteristic can be explained through so-called post-hoc approaches, i.e, generating explanations for contemplation after the training process~\cite{jentzsch2019don}. 
\paragraph{Explainee} The explainee, i.e. the recipient of the explanation, plays a central role; consequently, system’s explanations require a different level of detail for an expert or a developer compared to a naive, non-expert user. 

\subsubsection{Tools and approaches for XAI}
In the ML community, a row of explanation approaches has been established recently. These approaches mainly bear on ML models not intrinsically interpretable~\cite{molnar2020interpretable}, such as deep neural networks. \\%
Many XAI methods aim at highlighting the most relevant (to a certain outcome) features of the input data. In the case of neural networks, \textit{Layer-wise relevance propagation (LRP)}~\cite{montavon2019layer} works by propagating the prediction backward through the system and can be used to unmask correct predictions being made for the wrong reasons~\cite{lapuschkin2019unmasking}. Similarly, 
\textit{LIME} and \textit{SHAP} are python data visualization libraries. All mentioned approaches generate local post-hoc explanations and are model-agnostic.\\
A method partially related to unveiling correct decision being taken for false reasons are \textit{counterfactual explanations}, 
that determine and highlight which features need to be different to receive a different system outcome~\cite{vermacounterfactual}. \textit{Concept-based} explanations aim to identify relevant higher-level concepts instead of features specific to the input data. As such, they focus on meaningful human concepts, 
establishing human-understandable explanations~\cite{ghorbani2019towards,tcav_kim}. %
A non-technical measure to enhance the explainability and responsible deployment of intelligent systems is the convention of \textit{model cards}. This framework specifies relevant details regarding the model’s training, evaluation, and intended usage, which helps practitioners to understand the context and conclude assumptions about inner workings~\cite{mitchell2019model}. \\
The fusion of modern ML approaches with symbolic AI yields methods depicting learned representations from neural networks symbolically in an inherently intuitive structure.They appear highly effective for achieving interpretability, trust and reasoning (also see Sec.~\ref{s:results}). Primary methods of \textit{neural-symbolic learning}~\cite{nsvqa} aim at injecting semantics, as seen in~\cite{xai_semantic}, or expert knowledge in the form of knowledge graphs~\cite{xai_symbolic}.

\subsection{Human Perspective }\label{ss:background:human}
Following the preceding description of the technical perspective on the collaboration between humans and artificial intelligence, this section provides an overview of important concepts from human factors research on the collaboration and coordination between human operators and AI in domains where safety is critical. 

\subsubsection{Collaboration at work} 
\textit{Collaboration} is based on the human’s ability to participate with others in collaborative activities with shared goals and intentions, as well as the human’s need to share emotions, experiences, and activities with each other~\cite{tomasello2005search}. This enables people to work together and understand each other. As a consequence, human-centered integration of AI should address humans’ expectations on their human partners as well as digital partners. \\
In domains where safety is of critical importance and human error can have severe consequences~\cite{hauland2008measuring, salas2008teams}, human operators often work together in control centers~\cite{suchman1997centers} to achieve efficient and safe operation. Examples are airport operational centers, air traffic control centers, nuclear power plants, and military control centers. In control centers, teams of human operators have to work under time pressure to supervise complex dynamic processes as well as decide for remedy. Supervisory control is the human activity involved in initiating, monitoring, and adjusting processes in systems that are otherwise automatically controlled~\cite{national1930research}. Being a supervisor takes the operator out of the inner control loop for short periods or even for significantly longer periods, depending on the level at which the supervisor chooses to operate~\cite{wickens1997mavor}. %
Workshops with experienced pilots and ATCOs, which were conducted in order to gather their expectations about future tasks, roles, and responsibilities, indicated that task allocation, teamwork, and monitoring in a highly automated workplace pose challenges~\cite{bruder2008pilots}. As supervisory control is one of the core tasks in control rooms, teams of operators are required to monitor the systems appropriately~\cite{sharma2016eye}. Through interactions, operators in a team can dynamically modify each other’s perceptual and active capabilities~\cite{gorman2017measuring}. However, when monitoring a system, it is essential that human operators work together effectively and cooperatively~\cite{cooke2000measuring, salas2008teams}. %
With this in mind, communication in control operations is of high importance. Communication as a “meta-teamwork process that enables the other processes”~\cite{papenfuss2013phenotypes} provides indications for the coordinative activities while monitoring.Especially in critical situations, “it is not only critical that teams correctly assess the state of the environment and take action, but how this is accomplished”~\cite{cooke2013interactive}. As a consequence, a team’s communication provides insight into how the team members deal with critical situations.

\subsubsection {Trust and acceptance}    
Especially in domains where safety is of critical importance, both \textit{trust} of the users in human-human interactions and trust in human-technology interactions is of vital importance~\cite{bonini2001atc}. Trust as a psychological concept is defined as a belief in the reliability, truth, ability, or strength of someone or something~\cite{american2020apa}. Trust influences interpersonal relationships and interaction and plays a fundamental role in decision-making~\cite{dunning2011understanding} and risk perception~\cite{earle2010trust}. Trust can be influenced by past experiences~\cite{chen2010impact}, communication~\cite{blobaum2016trust}, and behaviors.\\
Trust in automation can be conceptualized as a three-factor model consisting of the human trustee, the automated trustee, and the environment or context. In this model, qualities of the human (such as experience), work with qualities of the autonomous agent (such as form) in an environment that also influences the nature of the interaction. Since trust is constantly evolving, time itself is also a facet of trust in human-automation interactions. Measurement of trust is challenging because trust itself is a latent variable, and not directly observable.~\cite{krueger2021neurobiology}. To make the complexity of the concept more manageable, technical perspectives often consider it as the extent to which a human believes the AI’s outputs are correct and useful for achieving their current goals in the current situation~\cite{Tomsett2020-dm}.\\
Trust and \textit{acceptance} are related in that a person is more likely to accept something if they trust it, which is investigated for users’ trust in AI technologies by \cite{choung2022trust}. Trust can provide a sense of confidence and security, which can make it easier for a person to accept something. In addition to the concept of trust, human acceptance of technology plays an important role. Technology acceptance is the extent to which individuals are willing to use and adopt new technological innovations \cite{silva2015davis,davis1985technology}. It is a multi-dimensional concept that takes into account various factors that influence an individual's decision to use a particular technology. The concept of technology acceptance is rooted in the theory of reasoned action \cite{hale2002theory} and the theory of planned behavior \cite{ajzen1991theory}. Technology acceptance is influenced by a range of factors, including the perceived usefulness and perceived ease of use of the technology, social influence, trust, compatibility with existing technologies and practices, perceived risks, and anxiety about using the technology.\\ 
To summarize the human perspective, it can be stated that a successful integration of AI in control centers’ operations has to consider humans’ expectations on their human partners and digital partners. To address the humans’ expectations on their human partners and digital partners, AI systems should:(1) support teamwork in in safety critical situations, (2) facilitate situation awareness, (3) consider the requirements of supervisory control, (4) support communication between team members, and (5) consider both interpersonal trust and trust in technology.

\subsection{Human-in-the-loop Methods} \label{ss:background:HITL}
\begin{figure}[!ht]
\centering
\includegraphics[scale=.52]{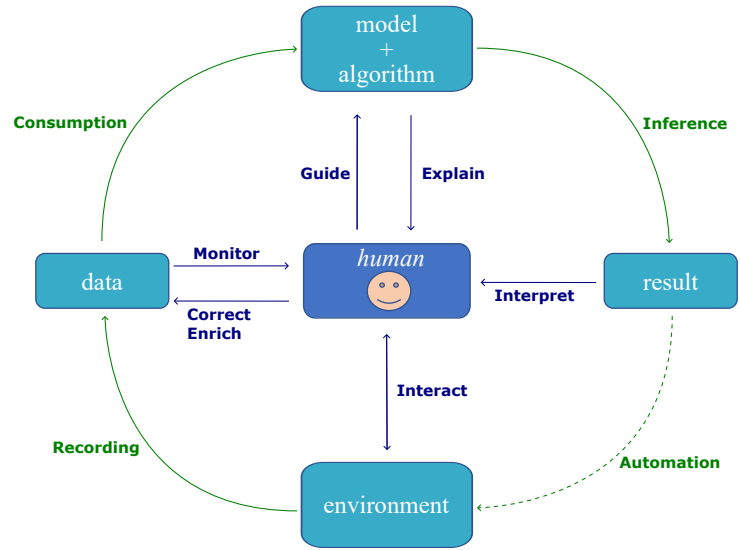}
\caption{
A figure depicting the data/information/knowledge/command flow -- showed as directional arrows -- in a classical \textit{human-in-the-loop} approach.
The AI-based system is abstractly depicted as consisting of the data -- recorded from the environment -- given as training/inference input; the model architecture and learning algorithm; followed lastly by its results (dependent on the AI’s task). The results might be used as commands for an automated system to act on the environment.
The blue arrows should be interpreted as the human - either a developer or a non-expert user - performing the action (e.g. “the human guides the model”, or “the human interprets the result”).
} 
\label{fig:hitl_overview}
\end{figure}

\newcommand{\hitl}{HITL}
The information flow in AI-based automated systems can be represented as a loop: the \textit{environment} is recorded using sensors; the \textit{data} produced by the recordings is consumed by the \textit{algorithm} to either train a \textit{model} or to use the model to infer a \textit{result}; the result is used as a command for an automation to modify the environment. In order to trust the system in safety critical applications, humans must have the oversight and understanding of the various elements of the loop. It is therefore natural to place the human in the loop (also see Fig.~\ref{fig:hitl_overview}).


\subsubsection{Definition}
The \emph{Human-in-the-loop} (\hitl) paradigm is a set of human oversight mechanisms on systems running AI models. Such mechanisms implement human-computer interaction methods at different levels of the AI-based system life-cycle such as data collection, model design, training process, model evaluation or model inference~\cite{altai, hitl_survey_wu, understand_hitl_cui}. Overall, \hitl{} brings together research fields from computer science, cognitive science and psychology~\cite{hitl_survey_wu}.


\subsubsection{Approaches}
The implementation of \hitl{} in a ML system primarily depends on the degree and nature of human knowledge to be injected. This can take place throughout the entire ML pipeline as seen in Fig.~\ref{fig:hitl_overview}. Following the categorization in~\cite{hitl_survey_wu} an initial approach is performing data processing with \hitl{}. The goal is obtaining a valid data set, i.e. which is accurately labeled (with the help of human annotators) essentially at key/representative data samples, stemming from a pool of unlabeled data. 
The above method employs expert feedback before the ML model training and inference take place. During training, feedback can be used to push the model to map its knowledge as closely as possible to humans’ (i.e, rewarding alignment with their decisions) or by learning through imitation~\cite{understand_hitl_cui} in the case of \textit{reinforcement learning (RL)} agents. 
Finally, \hitl{} coupled with model inference is best described by the application areas where the outcomes of ML models are used or processed by humans (occasionally interactively and iteratively). Collaboration is mostly imagined in this setting since the human has multiple abilities to interact with the outputs such as choosing between or observing multiple outcomes, to ask for explanations on what they represent or to further refine them by accepting/rejecting the AI’s assistive input~\cite{Assael.2022}.

\subsubsection{Applications} 
A \hitl{} system shows great results in domains where the human creativity and fine understanding of the context is combined with the machine’s data-analysis to reach performance superior to the human alone or the machine alone~\cite{Assael.2022}. %
Trust and acceptance can be built as well, as seen in~\cite{confl_res_phu} where \hitl{} for data labeling is employed to improve automatic conflict resolution suggestions within the air traffic management. Specifically, claiming that modern methods are not fully trusted by human operators, such as ATCOs and pilots, the authors enhance their acceptance by combining human-generated resolutions with RL algorithms.\\ 
The collaboration of human and AI-based systems -- as a hybrid team -- is particularly relevant in safety critical applications where the strengths of the machine on data-analysis and tasks repetitiveness are combined to the context understanding and adaptability to new scenarios of the human operator. As in a \textit{4-Eyes Principle} team organization, it is expected for the hybrid team to be less prone to missing relevant information or to overlooking effective solutions. On top of that, in collaborative RL schemes the safety of the human teaching the AI-based system is naturally prioritized. Consequently, random and/or dangerous actions of this system can be mitigated by sophisticated on-the-fly guidance from humans, as noted in~\cite{safe_hitl}.\\
Concerning the current and future focus in \hitl{} systems, the authors in~\cite{hitl_survey_wu} indicate that existing methods need to learn more effectively from expert human experience, essentially by moving towards more complex and less simplistic and superficial human intervention.

\section{Problem and Research Questions}\label{s:research_q}
XAI techniques provide information that describes how ML or DL models generate results based on data or processes. Depending on the model type, its results can be explained either by reducing complexity or by looking back at its training or evaluation. These techniques aim to demonstrate the effectiveness of models for developers of ML and DL models. However, to what extent information resulting from XAI methods aligns with the users’ requirements remains unclear. Although HITL approaches allow for human input, their purpose rather addresses improving the life cycle of the AI model, including its data collection, model design, training, and evaluation. So far, less attention is given to explanations serving the user and contextual goals. %
Human-computer interaction and user-centered design have long addressed the challenges of developing technical systems that meet user needs. Eliciting the requirements of different user groups may provide valuable insights for developing accepted and trusted~AI. In the run-up to requirements analysis, the following research question arises:
\begin{itemize}[itemsep=1ex, leftmargin=1.2cm]
    \item[\textbf{RQ1}:] What information do people need to perceive an AI system as explainable or understandable? 
    \item[\textbf{RQ2}:] What information do people need to accept an AI system?
    \item[\textbf{RQ3}:] Which interaction/ information representation methods are trustworthy?
\end{itemize}

\section{Method}\label{s:method}
In order to answer the aforementioned research questions, a structured literature review was conducted. Its methodological procedure is described in the following section, following the general methodological framework of the \textit{Preferred Reporting Items for Systematic reviews and Meta-Analyses (PRISMA)} statement published in 2009~\cite{moher2009preferred} and updated in 2020~\cite{page2021prisma}.

\subsection{Databases and Search Query}
Accordingly, a comprehensive literature search was conducted in the “Web of Science” and “Google Scholar” databases, the DLR repository “eLib,” the “DLR Library Catalog”, the “NASA Technical Report Server”, as well as the “Ebook Central” portal, and the database of German national libraries. We identified search terms and used Boolean operators to generate the following query strings for searching each of the mentioned sources the search term combination ((“explainability” OR “traceability” OR “acceptance”) AND (“artificial intelligence”) AND (“reasoning” OR “problem solving” OR “knowledge representation” OR “automatic planning” OR “automatic scheduling” OR “machine perception” OR “computer vision” OR “robotics” OR “affective computing”)).

\subsection{Identification and Screening}
Here, relevant articles in mentioned data sources were identified from 01.08.22 to 08.10.22. In total, $n = 244$ articles identified as relevant were returned from the “Web of Science Core Collection”, $n = 240$ relevant articles from a total of $n = 18,700$ Google scholar results, and $n = 27$ from the DLR search consisting of $n = 16$ from NASA Technical Reports, $n = 5$ from eLib publications, and $n = 6$ articles originating from the DLR library catalog. The latter were not included because source details were not available. %
The search results of all three queries were saved to .ris files and imported into the browser-based literature management program Paperpile, and data source tags were added here. When exporting the Web of Science Core Collection results, there was a loss of $n = 10$ records that were presumed duplicates. In addition, another record was recognized as a duplicate in Paperpile and removed from the initial records set. This resulted in $n = 521$ initially identified records as input for screening.
\subsection{Inclusion and Exclusion Criteria}
During the screening, each reference was screened first by machine filtering and then by manually checking the titles and abstracts for the following a priori inclusion and exclusion criteria: (1) the language of the article is in English, (2) the publication date of the article is between the years 1950 and 2022 inclusive, (3) the article contains results on explainability, and user acceptance of systems where humans interact and collaborate with AI, (4) the AI systems perform at least one of the following tasks: reasoning, problem-solving, knowledge representation, automated planning, and scheduling, ML, natural language processing , machine perception (computer vision), machine motion and manipulation (robotics), or emotional or social intelligence (affective computing). After excluding $n = 412$, banned records $n = 109$ could be retrieved and assessed in detail for mentioned inclusion and exclusion criteria. 
\subsection{Content Assessment}
In the subsequent systematic review of $n = 48$ reports, two different content assessments were made: (a) the implicit or explicit perspective of humans and (b) the quality of evidence for the outcome of interest. In addition, qualitative narrative analysis and synthesis through tabulation were performed. Of the 48 articles subjected to close examination, 32 were journals, 13 were conference papers, and 3 were technical reports—articles from 1980--2022. While early articles focus on describing technical implementation and function, the proportion of empirical evaluation of technical systems and inclusion of the user perspective increases over time.

\section{Results}\label{s:results}
The following section presents the key findings of previously described literature review, highlighting the information needed for explainability and acceptance together with trustworthy methods for humans interaction with AI systems, as well as trustworthy information representation methods.
\subsection{Information Needed for Explainability} \label{ss:results:info_needed_expl}
In this section, we explore the information needs of different users in understanding the results and behavior of AI systems. Analysis additionally revealed characteristics of explanations that are most effective in supporting human reasoning.

\subsubsection{Information explaining the model}
A much-regarded user group in the development of explainable AI are its developers, who have extensive technical and AI-specific background knowledge~\cite{Garibaldi2019-ei, Prentzas2019-jd, Beno2019-sh,Beyret2019-kq, Ieee2020-yb, De2020-vo}. For AI development, developers require information to understand the data and the model in terms of its internal operations, such as the weighting of individual parameters, features, or nodes, as well as information about the relation between input and output variables~\cite{Dam2018-dr, jentzsch2019don}. Such model-specific information can largely be generated through model-specific, global, and local XAI methods, as described in the background section of this paper. In addition, the contextual information of the use case or the development process may be important. However, these are rarely addressed by common XAI methods~\cite{Dam2018-dr, jentzsch2019don, Vorm2018-rn} and require at least HITL approaches. While model-specific XAI can enhance the explainability of AI techniques to developers~\cite{Spreeuwenberg2020-mq, Gilpin2018-zf,Gerdes2019-lq}, these effects do not necessarily carry over to non-expert end users we subsequently shed light onto their requirements.
However, explaining an AI through accessible raw data, code, or details about the AI models can also entail disadvantages, such as code manipulation and restrictions on the inventor’s potential for innovation. In this regard, a balance needs to be found between the need for transparency and the demand for ownership~\cite{Umbrello2022-dc}. \cite{Umbrello2022-dc} argue in their position paper to put more effort into understanding the requirements of all relevant user groups of an AI system to ensure that information for explaining the AI can be understood and thereby increase efficiency and effectiveness of the system.

\subsubsection{Information explaining results}
Most important information non-expert users require is information explaining the results as output or behavior of an AI system~\cite{Klumpp2019-kn,Dominick1984-mf, Simpson1999-tn, Zarka2016-yp, Ibrahim2020-nk} by answering the “WHY?”-question. This is often addressed by providing the raw data from which the results were derived from~\cite{Vorm2018-rn,Simpson1999-tn, Atkinson1991-ny,Murphy2004-xy, Shin2021-gz, Braun2021-ip} as well as through access to additional data used to generate the results, such as user interaction data~\cite{Zarka2016-yp}. 
Beyond that, users require information that explains why specific properties are assigned to certain result~\cite{Dam2018-dr,Alshammari2019-ax, Ibrahim2020-nk, Joshi2021-rf} and to which extent features~\cite{Winkler2017-vm, Dam2018-dr}, rules and decisions contributed to a specific result~\cite{Calvaresi2019-po}.
As with technical users, non-technical users need model-specific statistical information, data sources, and their biases or quality in terms of six dimensions, which are completeness, uniqueness, timeliness, validity, accuracy, and consistency~\cite{Joshi2021-rf, Burkart2021-cc}, and algorithmic information~\cite{Ibrahim2020-nk}.

\subsubsection{Characteristics of understandable explanations}
Information that explains the results or actions of an AI is particularly effective for humans if it supports logical inductive and deductive reasoning~\cite{Zarka2016-yp, Alshammari2019-ax, Vorm2018-rn, Ibrahim2020-nk}. Humans generally understand explanatory information best if they are presented in a contrasting manner~\cite{Vassiliades2021-ar, Joshi2021-rf,Burkart2021-cc}. Thereby, different properties of a result, different results with other properties, and results with different properties at different points in time should be compared to each other~\cite{Dam2018-dr}.\\
However, most model-agnostic, local XAI methods provide far more detail than what an end-user requires for a satisfactory explanation~\cite{Calvaresi2019-po}. Explanatory information should therefore include various levels of detail represented conditionally to context, explanation capability of the AI, and temporal, perceptual, and cognitive resources of the user.\\
Contextual information, domain knowledge, and meta-information such as time and location are vital for domain experts, e.g., in healthcare~\cite{Gaur2021-hq}. Regarding security-critical scenarios, information should be communicated at conflict-free and less work-intensive times~\cite{Klumpp2019-kn}. When representing model-specific statistics often discrete in nature, it has to be considered that human understanding and explanation of phenomena invariably utilize categories together with relationships among them, describing, for example, the relation between model predictors and the target as the relationship between entities and the target~\cite{Lukyanenko2020-ro}. Such relationships might contain probabilistic information even if these are not as crucial for humans as causal links~\cite{Vassiliades2021-ar, Burkart2021-cc}. Recent work even points out that most humans struggle to deal with uncertain information~\cite{Burkart2021-cc}. Causal information, in turn, supports humans, especially in decision-making in unfamiliar situations, but it has to be considered when individuals have prior experience with a domain, causal information can reduce confidence and lead to less accurate decisions~\cite{Zheng2021-vi, Burkart2021-cc}.

As humans prefer rare events, explanations should focus on odd reasons and be concise, meaning that shorter explanations are not considered interpretive. Form and explanation content interact largely with what is understandable~\cite{Burkart2021-cc}. To make it even more challenging, relevant contextual information extends to the person’s social context, considering assumptions about the users’ beliefs about themselves and their environment~\cite{Burkart2021-cc}. When including contextual information in an explanation, different users and situations have to be considered~\cite{Kastner2021-ok}.
\subsubsection{Application-specific information needs}
Since 2019, consideration of the user perspective has been increasing in the development of XAI. The outcome of a conversational agent supporting criminal investigations, for example, revealed that investigators want to have a clear understanding of the data, system processes, and constraints to make informed decisions and continue the investigation effectively~\cite{Hepenstal2021-tn}.\\
Furthermore, different user perspectives of autonomous surface vehicles (ASV) included AI developers, engineers, and expert users, who required information about the ASV models and training data used. Operators, crew, and safety attendants wanted to get information about the current state and intention of the ASV, as well as the definition of the AI-human control boundary and when to intervene. Passengers instead needed confirmation that the ASV can see and avoid collisions with other objects~\cite{Veitch2021-no}. \\
Last but not least, domain experts’ and lay users’ trust in a robotic AI system increased by providing relevant reasons for each of its decision together with explanations of the systems’ autonomous policy and the underlying reinforcement learning model through natural language question-answer dialogue~\cite{Iucci2021-sx}. \\

\subsection{Information Needed for Acceptance} \label{ss:results:info_needed_accept}
To increase the chances of humans accepting an AI, it is essential to understand what information they require for acceptance. The acceptance of artificial intelligence (AI) systems is considered a proxy measure for trust~\cite{Vorm2018-rn}, but can also emerge as a barrier to it~\cite{Veitch2021-no}. Further details about the relation of both concepts are described in the Background section.
\subsubsection{Goal-supporting information}
Literature suggests that information for acceptance strongly relates to the system’s functions and performance, demonstrating and emphasizing the use so that the perceived usefulness of an AI system increases~\cite{Klumpp2019-kn, Ismatullaev2022-lj}. Information supporting usefulness of a system is the goal-supporting information needed to successfully complete tasks contributing to the user’s goal. For example, if an AI as part of a guidance and control system for a spacecraft provides erroneous information about system states, key functions cannot be performed, resulting in direct user rejection of the system~\cite{Kraiss1980-ia}. In case of an AI recommending health decisions, correct general medical and patient-specific information together with best practice procedures are required~\cite{Simpson1999-tn} while for the acceptance of air traffic control systems, goal-supporting information include route information, air traffic information, sequential position and velocity information of other vehicles, clearance, events, vehicle responses, altitude, position of own vehicle, positions of other aircraft or information for contingencies, such as diagnosis of vehicle subsystems~\cite{Lowry2018-rm}. As with information for explainability, goal-supporting information strongly depends on the domain, task, and context of an AI system~\cite{Sousa1999-pj, Simpson1999-tn}. However, it has been shown that across application domains, the acceptance of an AI system can be increased by making the scope and limitations of AI methods and information about potential system failures~\cite{Kraiss1980-ia} known to users beforehand~\cite{Day1997-mk, Sousa1999-pj}.
\subsubsection{Reliability information}
The acceptance of AI results benefits from attached quality or reliability indicators such as error margins, uncertainties or confidence intervals, especially in high-stakes contexts~\cite{Vorm2018-rn,Sousa1999-pj}. All information provided must be tailored to their individual preferences and should, in general, include how data about the user is collected and processed, and how privacy is protected~\cite{Zarka2016-yp, Vorm2018-rn}. In particular, information about the extent to which other users trust the system plays a vital role in positively influencing its acceptance, especially if these are actors having a high social significance for the user, such as friends, family, work colleagues, or professional experts~\cite{Vorm2018-rn}. Arguments, for example applied to explain an AI and its results or actions~\cite{Vassiliades2021-ar}, are accepted if they have the support that makes them acceptable to the participants in a conversation. Similarly, information that establishes perceived usefulness and ease of use is related to user acceptance~\cite{Shin2021-gz}. %
The greater the coherence of a proposition with other propositions the greater its acceptability. If a proposition is highly coherent with a person’s beliefs, then the person will believe the proposition with a high degree of confidence and the other way around, also known as confirmation bias~\cite{Burkart2021-cc, Thagard1991-ud}.

\subsection{Information Representations and Interaction Methods}\label{ss:results:info_representation}
Methods for representing information and interacting with those in the context of human-AI collaboration are trusted if they exhibit a certain anthropomorphism such as natural language and speech~\cite{Dominick1984-mf, Dam2018-dr, Baclawski2020-xh, Tomsett2020-dm,Vassiliades2021-ar,Gaur2021-hq,Hepenstal2021-tn,Pierrard2021-zz}, text~\cite{Burkart2021-cc, Sachan2020-qq} or human like visual appearances, for example in the context of robotics~\cite{Ene2019-zm, Goodman2008-hj, Shin2021-gz}.
\subsubsection{Textual and speech representations}
The former include text-based, as well as speech-based input and output. Especially domain experts expect system feedback in natural language to domain specific language~\cite{Dominick1984-mf} and be presented within 3-4 seconds~\cite{Goodman2008-hj, Lowry2018-rm} to ensure a cognitive and emotional linkage through realistic, social interaction. Social quality of a dialogue through emotionally intelligent interaction can profit from closed-loop interaction with cognitive human models~\cite{Goodman2008-hj}.In order to address the previously described information need by answering "why” questions, but also “how” and “why not” questions, natural language should be expressed easily understandable in a narrative style~\cite{Baclawski2020-xh}.Dialectic explanations have for example been generated based on a log file with internal steps an AI performed to reach a certain recommendation~\cite{Vassiliades2021-ar}.
\subsubsection{Data and information visualizations}
In addition to natural language and speech interaction, data and information visualizations such as graphs~\cite{Atkinson1991-ny, Winkler2017-vm,Hepenstal2021-tn}, charts~\cite{Murphy2004-xy}, and animations~\cite{Burkart2021-cc} are especially suitable for efficiently conveying information from statistical~\cite{munzner2014visualization} and model-specific data~\cite{Vassiliades2021-ar, Gaur2021-hq} such as intermediate network layers of DNNs~\cite{Winkler2017-vm}, neuron activation and weights or token embedding in 2D and gradient based methods~\cite{Joshi2021-rf}. In addition, visualizations suitable for the representation of structural information such as CNN feature maps or DNNs graph structures~\cite{Dam2018-dr, Joshi2021-rf} or conceptual and semantic information~\cite{Gaur2021-hq, Joshi2021-rf}. Features impact such as words impact on the classification outcome could effectively be represented through
color-coding~\cite{Winkler2017-vm}, especially when coding is based on relations relevant grammars~\cite{Lukyanenko2020-ro}. Even though, speech is frequently used to represent explanatory information and multi-modal data contains persistent inconsistencies and biases~\cite{Joshi2021-rf}, combining graphic narratives with natural language can be even more effective~\cite{Baclawski2020-xh, Vassiliades2021-ar, Yokoi2021-ts, Shin2021-gz, Murphy2004-xy}, reduce human workload or increase human performance~\cite{wickens2021engineering}. Visual representations are particularly suitable for target groups with little background knowledge; analogies that correspond to the mental model can reinforce this~\cite{Veitch2021-no}.
\subsubsection{Interaction quality}
Regardless of modality, safety-critical contexts often require interactive and reciprocal information exchanges and learning among humans and machines (HITL)~\cite{Kraiss1980-ia, Joshi2021-rf}, while answers are expected to be fast and accurate~\cite{Zarka2016-yp}. Depending on user task and context touch and gesture-based interaction methods have also demonstrated to be powerful and effective~\cite{Klumpp2019-kn} while emotion-aware mechanisms, especially when combined with human-like appearance support user satisfaction and adherence\cite{Shin2021-gz}. In any case suitability of a representation or interaction method is strongly dependent on age, culture and gender of the user group~\cite{Ismatullaev2022-lj}. Examples for effective and efficient information representations and interaction methods include logged interactions to handle lost link procedures and error-free resumption of interaction after interrupted communication for an an artificial pilot. This system applied natural language interaction to interact with the terminal crew, to enable automated reasoning and decision making, to coordinate autonomous operations and basic pilot procedures with variable autonomy~\cite{Lowry2018-rm}.

\section{Discussion and Conclusion}\label{s:discussion}
This article aims to bridge the gap between technical and human perspectives in developing AI systems that are understandable, acceptable, and trustworthy. To achieve this, user needs are identified and transformed into requirements for AI system design, constituting an initial step for requirements engineering. These requirements must be validated and refined for various application domains to serve as the foundation for development activities.
\subsection{Contribution}
The results show that the existing methods for explaining AI (see Sec.~\ref{ss:background:technical}) correspond to the needs and requirements of people with extensive background knowledge about AI, ML and DL models and whose task is to develop and improve AI models. In contrast, people who have little AI background knowledge use an AI system to achieve their individual goals and to process application-related tasks. They mainly expect the results and behavior of the system to be explained. Only occasionally the latter group would like to use the statistical parameters of an AI model to understand the system result or behavior. However, for this purpose, a lower and more flexible level of detail is required than for the former group. Relevant to either group yet is the questionable reliance of certain XAI methods, with many being prone to manipulation and adversarial attacks \cite{slack2020fooling}. Arguably, multiple well-established explanation algorithms are criticized in~\cite{sanity_adebayo} revealing that some fail to depict accurate mappings from input features to model outcomes~\cite{fragile_ghorbani,unreliable_kindermans,slack2020fooling}. To further provide explanations that fit the user needs, their requirements have to be taken into account during the development but also within AI applications (as described in Sec. \ref{ss:background:HITL}). Since explanations should maximize the user’s mental model of explanatory information, human feedback should be incorporated to a greater extent to iteratively improve development outcomes and AI result. As an example, such an approach was followed in~\cite{ghorbani2019towards} and appears highly promising. %

A particular user group for AI systems collaborating with humans, are domain experts, such as ATCOs, medical professionals, or scientific personnel as crew and operators of space systems. They have extensive domain knowledge but restricted background knowledge of AI technologies. They also mainly need information that explains system results and behavior, aiming to understand a system outcome and behavior by information of the professional context rather than the technology. For medical professionals, this means, for example, that they want to interpret a result based on its relevance for different patient groups or based on its validity and relevance for other experts. Regarding the requirements for an AI system for air traffic control in terms of explainability, it can be stated that the needs of AI developers, as well as non-expert users, have to be considered: The former require information about the data and the model in terms of internal operations and the relationship between input and output variables. Essentially, the latter profit from information about the results and behavior of the AI system, why certain properties are assigned to specific results, and the contribution of features, rules, and decisions to a specific result. In general, the information should be presented in a contrasting manner and include various levels of detail depending on the context of each user group, the explanation capability of the AI, and the temporal, perceptual, and cognitive resources of individual users. For security-critical scenarios such as air traffic control, information should be communicated at conflict-free and less work-intensive times. Designing a comprehensible AI system requires various functionalities and modules that detail the individual needs and characteristics of all important user groups.

With respect to the information users require to accept an AI system, it can be stated that acceptance for AI systems profits from task-related information supporting users in achieving their goals, information demonstrating the performance and usefulness of the AI and information about privacy and ethical considerations. It has to be stated that information alone are not sufficient for acceptance the system in general needs to be useful in a user-friendly way also providing control over their data and data processing.\\
Regarding trustworthy information representation and interaction methods, results revealed that natural language and visual information representations are most suitable for human AI-collaboration, especially their combination. Effective interaction in safety-critical contexts, such as air traffic control, primarily require fast and accurate information exchange and learning between humans and machines. However, the suitability of a representation or interaction method is dependent on factors such as task and its context, the user’s age, culture, and gender.
\subsection{Limitations}
Results presented describe user needs and requirements for a system only to the extent that these were included in the literature. In this context, the underlying data is subject to time-dependent biases towards technology-centered development methods, limited result validity, and biases due to the topicality of applied models. Early work, for example, developed models with much smaller data sets which is why the users’ need to access these data might be more valid for earlier than for present systems. %
Literature analysis and synthesis was guided by the three research questions formulated in Sec.~\ref{s:research_q}. In order to provide the most comprehensive and generally valid information possible, no restrictions were placed on the fields of application, user groups, technologies, or research methods/questions. Accordingly, considered papers exhibit a high level of heterogeneity regarding these characteristics. Nevertheless, contextual and methodological variance among studies examined must be acknowledged as a potential limitation. However, its effect is reduced by the fact that the user requirements formulated here are validated, refined, and supplemented for the air traffic control application context.
\subsection{Future Work}
This literature analysis demonstrated that, with respect to interdisciplinary perspectives in developing AI systems, different frameworks for considering humans are exploited which are not being integrated enough. On the one hand, the human-in-the-loop paradigm as a set of human oversight mechanisms on systems running AI models is well-known within the technical community. On the other hand, human factors specialists and psychologists have adopted a human-centered design approach, in a framework that develops socio-technical systems by involving the human perspective in all steps of the design process. Finally, in safety-critical contexts such as air traffic control, research and development focuses on shifting from manual control where the human is \textit{in} the loop, to supervisory control where the human operator is \textit{on} the loop. By having integrated automation in aviation some decades ago, the human operator no longer needs to be in direct control of the system. As a result operators are supervising many aspects of the system, which changes the role of the human in a system.\\ 
Therefore, one main topic of future work is to share and integrate the different perspectives and methods for designing understandable, acceptable, and trustworthy AI systems in an interdisciplinary development team. Another topic for further research lies on investigating and validating the presented findings for AI integration in air traffic control systems. A first step is to conduct user workshops as to assess their expectations on tasks to be allocated between human and AI system, on information needed from AI systems, user-friendly interaction and use of personal data. In doing so, a two-day workshop with nine ATCOs from German air navigation service provider (DFS) and Austro Control GmbH is currently being conducted. Furthermore, it is planned to research and validate prototypes of AI systems with users throughout the design process of AI systems in aviation. To achieve this, experimental studies will be conducted in laboratory settings simulating a control-center task environment, as well as large-scale simulations of air traffic control with experienced operators. In doing so, guidelines for effective and safe collaboration between AI systems and human operators in safety-critical contexts will be investigated, which will finally lead to recommendations for the development of AI systems.

\bibliographystyle{splncs04}
\bibliography{references/main}
\end{document}